\documentclass[aps,prc,preprint,superscriptaddress]{revtex4-1}
\usepackage{graphicx}

\begin{document}

\title{Effect of isospin dependent cluster recognition on the observables in heavy ion collisions}

\author{Yingxun Zhang} 
\affiliation{China Institute of Atomic Energy, P.O. Box 275 (10), Beijing 102413, P.R. China}

\author{Zhuxia Li} 
\affiliation{China Institute of Atomic Energy, P.O. Box 275 (10), Beijing 102413, P.R. China}

\author{Chengshuang Zhou} 
\affiliation{China Institute of Atomic Energy, P.O. Box 275 (10), Beijing 102413, P.R. China}
\affiliation{College of Physics and Technology,Guangxi Normal University, Guilin 541004,P.R. China}

\author{M.B.Tsang}  
\affiliation{National Superconducting Cyclotron Laboratory and Joint Institute of Nuclear Astrophysics, Michigan State University, East Lansing, MI 48824, USA}
\date{\today}

\begin{abstract}
We introduce isospin dependence in the cluster recognition algorithms used in the Quantum Molecular Dynamics model to describe fragment formation in heavy ion collisions. This change reduces the yields of emitted nucleons and enhances the yields of fragments, especially heavier fragments. The enhancement of neutron-rich lighter fragments mainly occurs at mid-rapidity. Consequently, isospin dependent observables, such as isotope distributions, yield ratios of $n/p$, $t/^3He$, and isoscaling parameters are affected. We also investigate how equilibration in heavy ion collisions is affected by this change.
\end{abstract}


\maketitle

The productions of nucleons, light particles and intermediate mass fragments (IMF) in heavy ion collisions(HICs) provide unique experimental information about the equation of state(EoS)\cite{Brown91, Tsang09, Danie02, BALi08} as well as other equilibrium and non-equilibrium properties of nuclear matter far from its normal state \cite{Bondorf95, Gross90, Aichel91, Maria04, Chomaz04, Fevre08, Das05}. Interpretation of the experimental results requires comparisons of data to predictions from transport models which take into account reaction dynamics. Transport models track the time evolution of a nuclear reaction, and separately treat the nuclear EoS through the mean field part and the nucleon-nucleon (NN) collision cross sections through collision part. The influence of the mean field, both isospin dependent and independent parts, and the NN collision cross sections has been studied using both the Boltzmann-Uehling-Uhlenbeck (BUU) and Quantum Molecular Dynamics (QMD) approaches \cite{Zhang12, Coupl11, BALi05}. These studies also point out the importance of cluster formation in the description of reaction dynamics \cite{Zhang12,Tsang09, Coupl11, Ono03}.

In the QMD model, fragments are formed due to A-body correlations caused by the overlapping wave packets and are identified by the cluster recognition method which also plays important roles on the final observables. At any time of the reaction process, fragments can be recognized by a minimum spanning tree (MST) algorithm\cite{Aichel91,Zhang05}. In this algorithm, nucleons which have a neighbor within a distance of coordinate and momentum of $|r_i-r_j|\leq R_0$ and $|p_i-p_j|\leq P_0$ belong to a fragment. Here, the $r_i$ and $p_i$ are the centroids of the wavepacket for $i^{th}$ nucleon in their spatial and momentum space. $R_0$ and $P_0$ are phenomenological parameters determined by fitting the global experimental data, such as the intermediate mass fragments multiplicities \cite{Aichel91,Zhang05}. They should roughly be in the range of nucleon-nucleon interaction. Typical values of $R_0$ and $P_0$ used in the QMD approaches are about 3.5 fm and 250 $MeV/c$\cite{Aichel91, Zhang05}. This approach has been quite successful in explaining certain fragmentation observables, such as charge distributions of the emitted particles, intermediate mass fragments multiplicities \cite{Aichel91, Zhang05,Zhang06, Zhang07}, yield ratios of free neutron to proton ($n/p$), and the double $n/p$ ratios in heavy ion collisions \cite{Zhang08}. On the other hand, the MST method fails to describe other details in the production of nucleons and light charged particles \cite{Zhang05, Aichel91, Nebau99}. For example, while the yields of Z=1 particles are overestimated, the yields of Z=2 particles are underestimated partly due to the strong binding of $\alpha$ particles. Enhancements of the production of neutron-rich isotopes observed in isoscaling \cite{TXLiu}, dynamically emitted heavy fragments \cite{Russo} in neutron-rich HICs, and neutron-rich light charged particles (LCP) at mid-rapidity \cite{Kohley} have not been very described well by simulations using transport models. Furthermore, most transport models predict more transparency than that observed experimentally in central collisions at intermediate energy \cite{Zhang12, Hudan03} due to insufficient production of fragments in the mid-rapidity region. Previous studies show that these problems can not be resolved by only changing the mean field or nucleon-nucleon cross section in transport models.

There have been many attempts to improve the MST algorithm. More sophisticated algorithms such as the early cluster recognition algorithm (ECRA)\cite{Dorso93}, the simulated annealing clusterization algorithm (SACA) \cite{Goss97, Puri2000}, and the minimum spanning tree procedure with binding energy of fragments (MSTB) \cite{Goyal11} have been developed to provide better description of the IMF multiplicities or the average $Z_{max}$ of the fragments. However these algorithms do not address the lack of isospin dependence in cluster recognition which is the main focus of this paper.

In the regular MST method, neutrons and protons are treated equally in the cluster recognition process, a constant maximum distance for the nucleons to be defined as belonging to a cluster is $R_0^{nn}=R_0^{np}=R_0^{pp}=R_0\sim3.5 fm$. In order to investigate the effects of introducing the isospin dependence in the cluster recognition, the iso-MST, we choose different values of $R_0^{nn}$, $R_0^{np}$, and $R_0^{pp}$. We found that $R_0^{nn}=R_0^{np}=6fm$ and $R_0^{pp}=3fm$ give the largest effect on the heavy ion collisions observables in this study. These values are consistent with the $R_0$ values ranging from 3 to 6 fm used in the QMD simulations in order to reproduce the cluster distribution reasonably. The larger distances of $R_0^{nn}$ and $R_0^{np}$ take into consideration of the properties of neutron-rich nuclei, such as neutron skin or neutron halo effect. Due to the long range repulsive Coulomb force between nucleon in the cluster, a smaller $R_0^{pp}$ value is adopted. $P_o =250 MeV/c$ is the same in both the MST and the iso-MST methods because nucleons with large relative momentum are no longer close together in coordinate space after some time. In this paper, we compare the results calculated with the MST and the iso-MST methods to investigate effects of introducing isospin dependence in the identification of cluster formation on isospin sensitive observables. We also investigate how these cluster recognition methods affect the observable on system equilibrium. All of our studies are based on the ImQMD05 code \cite{Zhang05,Zhang06,Zhang07,Zhang08}.

Within the ImQMD05 approach, nucleons are represented by Gaussian wavepackets and the mean field acting on these wavepackets is derived from an energy functional with the potential energy $U$ that includes the full Skyrme potential energy without the spin-orbit term:
\begin{equation}
U=U_{\rho}+U_{md}+U_{coul}
\end{equation}
and $U_{coul}$ is the Coulomb energy. The nuclear contributions are represented in a local form with
\begin{equation}
 U_{\rho,md}=\int u_{\rho,md}d^3r
\end{equation}
and
\begin{eqnarray}
u_{\rho}&=&\frac{\alpha}{2}\frac{\rho^2}{\rho_0}+
\frac{\beta}{\eta+1}\frac{\rho^{\eta+1}}{\rho_0^{\eta}}+
\frac{g_{sur}}{2\rho_0}(\nabla\rho)^2\nonumber\\
&  &+\frac{g_{sur,iso}}{\rho_0}[\nabla(\rho_n-\rho_p)]^2\nonumber\\
&  &+\frac{C_s}{2}(\frac{\rho}{\rho_0})^{\gamma_i}\delta^2\rho
+g_{\rho\tau}\frac{\rho^{8/3}}{\rho_0^{5/3}}
\end{eqnarray}
Here, $\delta$ is the isospin asymmetry. $\delta=(\rho_n-\rho_p)/(\rho_n+\rho_p)$, $\rho_n$ and $\rho_p$ are the neutron and proton densities, respectively. A symmetry potential energy density of the form $\frac{C_s}{2}(\frac{\rho}{\rho_0})^{\gamma_i}\delta^2\rho$ is used in the following calculations. In the present work, we adopt the asy-soft value of $\gamma_i=0.5$ which provides the better agreement with data \cite{Zhang08, Sun10}. The energy density associated with the mean-field momentum dependence is represented by
\begin{eqnarray}
u_{md}&=&\frac{1}{2\rho_0}\Sigma_{N1}\frac{1}{16\pi^6}\int d^3p_1d^3p_2
f_{N1}(\vec{p_1})f_{N2}(\vec{p_2})\nonumber\\
&  &1.57[\ln(1+5\times10^{-4}(\Delta p)^2)]^2
\end{eqnarray}
where $f_N$ are nucleon Winger functions, $\Delta p=|\vec{p_1}-\vec{p_2}|$, the energy is in MeV and momentum is in MeV/c. The resulting interaction between wavepackets is described in Ref.\cite{Aichel91}. In this work, the $\alpha=-356 MeV$, $\beta=303 MeV$, $\eta=7/6$, and $g_{sur}=19.47 MeVfm^2$, $g_{sur,iso}=-11.35 MeVfm^2$, $C_s=35.19 MeV$, and $g_{\rho\tau}=0 MeV$. These calculations use isospin-dependent in-medium nucleon-nucleon scattering cross sections in the collision term and the Pauli blocking effects are described in \cite{Zhang05,Zhang06,Zhang07}.

\begin{figure}[htbp]
\centering
\includegraphics[angle=270,scale=0.5]{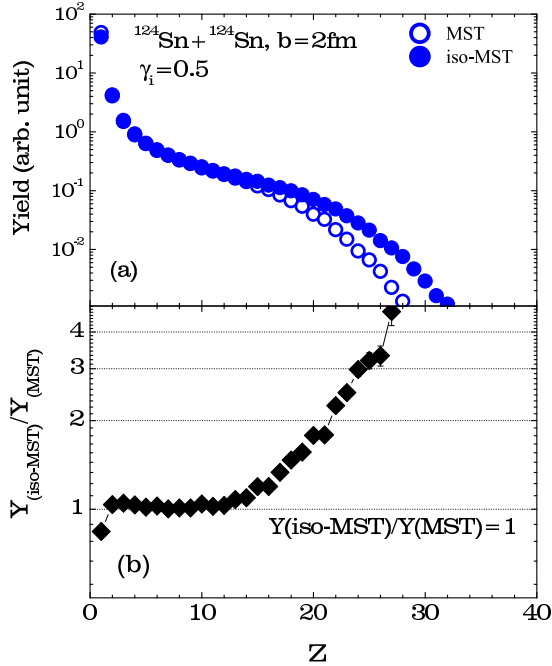}
\setlength{\abovecaptionskip}{50pt}
\caption{\label{ref-mc-dis-ratio-fig1}(Color online) (a) Charge distribution of $^{124}Sn+^{124}Sn$ at beam energy of $E/A=50 MeV$ for b=2fm for $\gamma_i=0.5$. The solid circles are results from the iso-MST algorithm and the the open circles for the same reaction system. (b)$Y(iso-MST)/Y(MST)$ ratio obtained from the same reaction system.}
\setlength{\belowcaptionskip}{0pt}
\end{figure}

Fig.1 (a) shows results of charge distributions from the central collisions of $^{124}Sn+^{124}Sn$ at b=2fm at 50 MeV per nucleon beam energy calculated in the MST (open circles) and the iso-MST (solid circles), respectively. In the current paper, we adopt the convention that open circles are results obtained with the MST method, and solid circles represent results using the iso-MST
method except in Figure 4. Fig.1(b) plots the yield ratios from the iso-MST and the MST algorithms, i.e., $Y(iso-MST)/Y(MST)$. Since larger values of $R_0^{nn}$ and $R_0^{np}$ are adopted, more nucleons, especially neutrons, are included in clusters. As a consequence, the multiplicities of Z=1 particles are reduced by about $16\%$, and the production of heavier fragments with $Z>12$ are strongly enhanced. This may explain the strong enhancement of dynamically emitted heavy fragments observed in neutron-rich heavy ion collisions \cite{Russo}. Even for intermediate mass fragments with Z=2-12, the multiplicities are larger by $\sim$3\%. In all cases, the enhancement is larger in neutron-rich heavy ion collisions.

In order to understand the enhancement of neutron-rich particles, we analyze the rapidity distributions of the particles emitted in central collisions of $^{124}Sn+^{124}Sn$ reactions at b=2fm and incident energy of $E/A=50 MeV$ using both the MST (open circles) and the iso-MST (solid circles) algorithms. In Fig.2, we present the rapidity distributions of \emph{n}, \emph{p},and their ratios in (a), (b) and (c), respectively. The calculations with the iso-MST method (solid circles) reduce the yields of both neutrons and protons over all rapidity regions relative to the results obtained with MST case, especially at mid-rapidity region. However, the yield ratios, $Y(n)/Y(p)$ (Figure 2 (c)) are enhanced at mid-rapidity and become smaller at forward and backward rapidity regions in the iso-MST case than that in the MST case.
\begin{figure}[htbp]
\centering
\includegraphics[angle=270,scale=0.5]{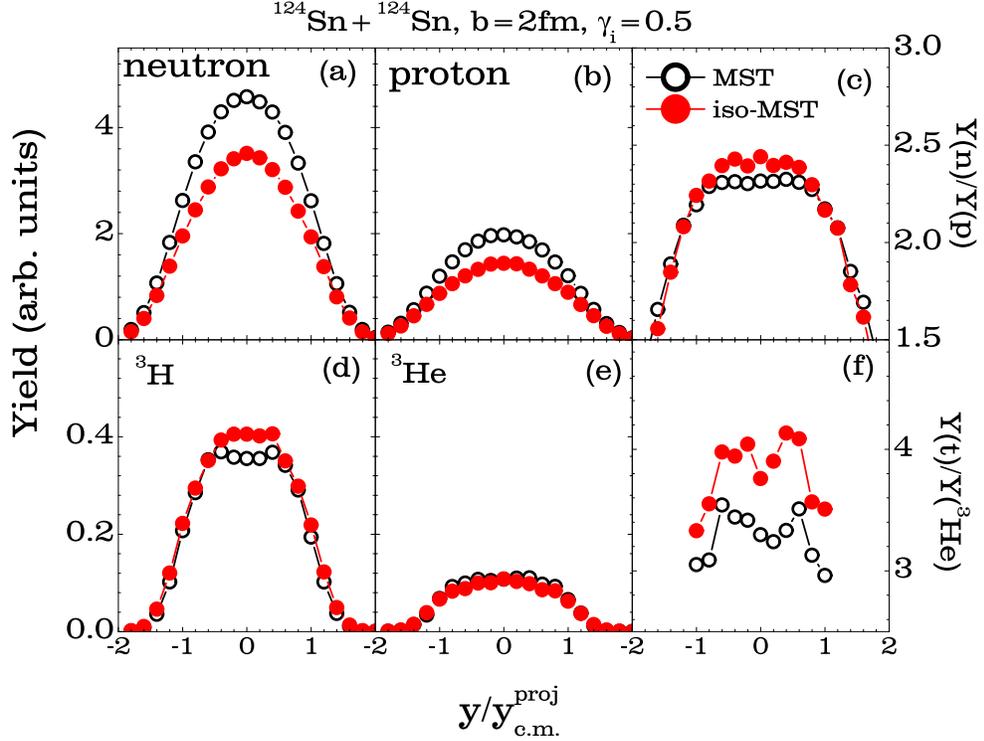}
\setlength{\abovecaptionskip}{50pt}
\caption{\label{ref-np-ratio-y-fig2}(Color online) The rapidity distribution of (a) the neutron yields, (b) the proton yields (c) the yield ratios, $Y(n)/Y(p)$. (d) the \emph{t} yields, (e) the $^3He$ yields and (f) the yield ratios, $Y(t)/Y(^3He)$.}
\setlength{\belowcaptionskip}{0pt}
\end{figure}

In Figure 2 (d), (e) and (f), we plot the rapidity distributions of $t$, $^3He$ and their ratios, $Y(t)/Y(^3He)$, respectively. The $t$ and $^3He$ yields are much lower than that of $n$ and $p$. Note that the scales of the y-axis for Fig. 2 (d) and (e) are a factor of 10 smaller than those in Figure 2(a) and (b). Unlike the neutron yields, the triton yields (Fig. 2d) obtained with the iso-MST method are enhanced at mid- rapidity, and remain nearly the same in the projectile or target regions relative to the results obtained with MST method. On the other hand, there is very little difference in the yields of proton-rich $^3He$ (Fig. 2e) obtained with the iso-MST and the MST methods over the whole rapidity regions. As a consequence, the yields ratios, $Y(t)/Y(^3He)$, are much enhanced as shown in Figure 2 (f). The values of $Y(t)/Y(^3He)$ obtained with the iso-MST method increase to about 4.0 at mid-rapidity, much higher than the $Y(n)/Y(p)$ ratios. Two factors contribute to the larger $Y(t)/Y(^3He)$ ratios at mid-rapidity. First, the emitted nucleons at mid- rapidity have lower kinetic energy, because most of them pass through the expansion phase and dissipate their kinetic energy. At the end of the simulations, the emitted nucleons with lower kinetic energy are closer to the other clusters. Thus, the iso-MST algorithm with larger $R_0^{nn}$ and $R_0^{np}$ values will reduce the yields of nucleons at mid-rapidity as more nucleons close to the clusters are absorbed into fragments. Furthermore, production of neutron-rich fragments is enhanced at mid-rapidity due to larger $R^{nn/np}_0$ values adopted in the iso-MST case. The calculated results in Fig.2 clearly demonstrate that cluster recognition method changes the yields of nucleons and light charged particles, and neutron-rich particles emitted at mid-rapidity in heavy ion collisions.

\begin{table}[b]
\caption{\label{tab:table1}%
The calculated results of R(n/p), R(t/$^3$He) obtained for $^{124}Sn+^{124}Sn$ at b=2fm, the results of DR(n/p), DR(t/$^{3}$He) and $\alpha$ are obtained with an angular gate of $70^\circ \le \theta_{cm}\le110^\circ$ from two reaction systems $^{124}$Sn+$^{124}$Sn, $^{112}$Sn+$^{112}$Sn at b=2fm.
}
\begin{ruledtabular}
\begin{tabular}{lcc}
\textrm{Observable\footnote{All the results are obtained with $\gamma_i=0.5$ case.}}&
\textrm{MST}&
\textrm{iso-MST}\\
\colrule
  $R(n/p)_{E_{k}<40MeV}(^{124}Sn)$ & $2.131\pm0.005$ & $2.302\pm0.006$ \\
  $R(n/p)_{E_{k}>40MeV}(^{124}Sn)$ & $1.044\pm0.011$ & $1.041\pm0.011$ \\
  $DR(n/p)_{E_k<40MeV}$ & $1.514\pm0.005$ & $1.641\pm0.006$ \\
  $DR(n/p)_{E_k>40MeV}$ & $1.803\pm0.029$ & $1.852\pm0.030$ \\
  $R(t/^3He)_{^{124}Sn}$ & $3.031\pm0.026$ & $3.497\pm0.028$ \\
  $DR(t/^3He)_{E_k/A<40MeV}$ & $1.57\pm0.02$ & $1.64\pm0.02$ \\
  Isoscaling parameter $\alpha$ & 0.19 & 0.22 \\
\end{tabular}
\end{ruledtabular}
\end{table}

We also analyze the influence of the isospin dependent cluster recognition method on isospin dependent observables, such as $R(n/p)=Y(n)/Y(p)$, $DR(n/p)$, $R(t/^3He)$ and $DR(t/^3He)$ at lower and higher kinetic energy regions. In Table I, we list the yield ratios of $n/p$ and $t/^3He$ with $E_k\leq40MeV$ and $E_k\geq40MeV$ for the neutron rich reaction system $^{124}Sn+^{124}Sn$. It is clear that the iso-MST case increases the values of $R(n/p)$ at lower kinetic energy relative to the MST case. To reduce systematic uncertainties in experiments, we also construct the double ratios $DR(n/p)$ obtained from two reaction systems $^{124}Sn+^{124}Sn$, $^{112}Sn+^{112}Sn$ at b=2fm using both the iso-MST and the MST algorithms. Here
\begin{equation}
    DR(n/p)=\frac{[Y(n)/Y(p)]_{^{124}Sn+^{124}Sn}}{[Y(n)/Y(p)]_{^{112}Sn+^{112}Sn}}
\end{equation}
For nucleons with low kinetic energy ($<40 MeV$), $DR(n/p)$ obtained in the iso-MST case is larger by about 8.3\%. For nucleons with high kinetic energy ($>40 MeV$), the difference between the $DR(n/p)$ values obtained from both the MST and the iso-MST cases is much less. As most of the emitted triton and $^{3}He$ have kinetic energies less than 40 MeV per nucleon, we list the $R(t/^3He)$ and $DR(t/^3He)$ values with $E_k/A<40 MeV$ in the fifth and sixth row of Table I. The iso-MST algorithm enhances the $R(t/^3He)$ and $DR(t/^3He)$  values more than $R(n/p)$ and $DR(n/p)$.

\begin{figure}[htbp]
\centering
\includegraphics[angle=270,scale=0.5]{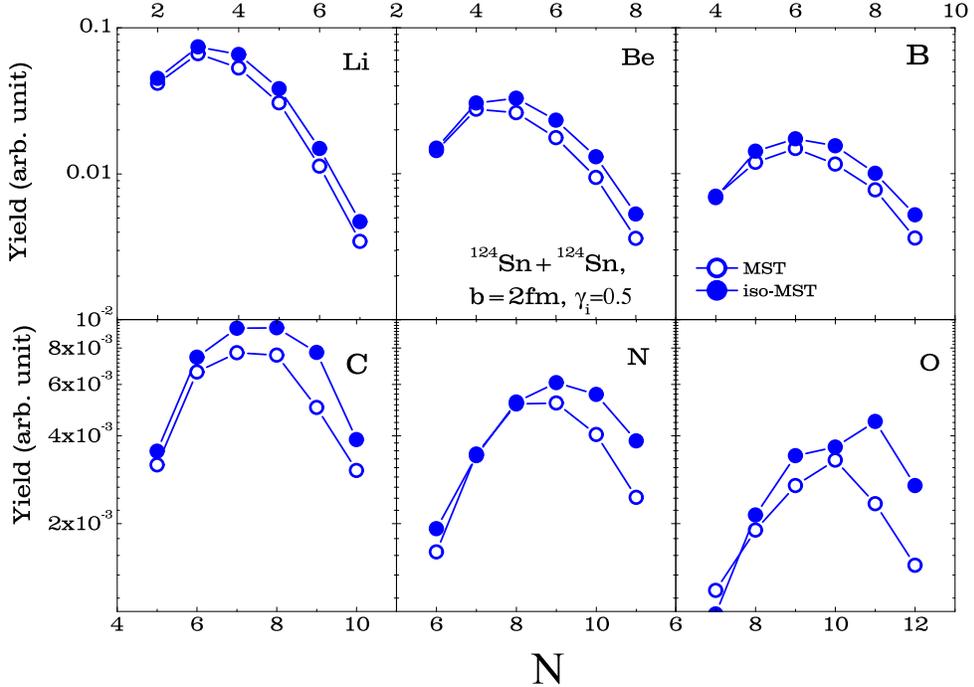}
\setlength{\abovecaptionskip}{50pt}
\caption{\label{ref-isodis-fig3} (Color online) Calculated primary isotope yields of \emph{Li}, \emph{Be}, \emph{B},
 \emph{C}, \emph{N} and \emph{O} for $^{124}Sn+^{124}Sn$ with an angular gate of $70^{\circ}\leq \theta_{c.m.}\leq 110^{\circ}$
 for $\gamma_i=0.5$ at b=2fm.}
\setlength{\belowcaptionskip}{0pt}
\end{figure}

With larger $R_0^{nn}$ and $R_0^{np}$ values adopted in the iso-MST method, more neutron-rich isotopes are produced. This effect is further enhanced in the neutron-rich reaction system. Fig.3 shows the isotope distributions of primary fragments with Z=3 to 8 predicted by the ImQMD05 calculations over the angle region, $70^{\circ}\leq \theta \leq 110^{\circ}$ for $^{124}Sn+^{124}Sn$ system. The solid circles are results obtained with the iso-MST method, and open circles are results obtained with the MST method.

The isoscaling relationship, which is constructed using isotope yields $Y_i(N,Z)$ with neutron number \emph{N} and proton number \emph{Z} from two different reaction systems denoted by the index \emph{i}, obeys a simple relationship
\begin{equation}
    R_{21}(N,Z)=Y_2(N,Z)/Y_1(N,Z)=Cexp(\alpha N+\beta Z)
\end{equation}
Here, $C$ is an overall normalized factor and $\alpha$ and $\beta$ are isoscaling parameters. The isoscaling parameter $\alpha$ can be used to estimate the enhancement of neutron rich isotopes, and has been used as a probe to study the density dependence of symmetry energy \cite{HSXu, Tsang01, Souza09, TXLiu}. We analyze the isoscaling relationship with both cluster recognition methods for fragments with $Z=3-8$. In Fig.4, as an example, we plot the $R_{21}=Y_2/Y_1$ values as a function of N of the emitted fragments obtained with the iso-MST and the MST cases. Here, "2" represents the neutron-rich reaction system $^{124}Sn+^{124}Sn$, and "1" represents $^{112}Sn+^{112}Sn$. The upper panel shows the results obtained with the iso-MST case and the lower panel shows results obtained with the MST case. As in previous study \cite{Ono03}, we find that the isoscaling relationship exists in non-equilibrium transport model. The best fit isoscaling parameters $\alpha$ are listed on the figure and in Table.I. The isoscaling parameter $\alpha$ obtained with the MST is 0.19, and the $\alpha$ value obtained with the iso-MST case increases to 0.22 which is still less than the experimental value of $\alpha=0.36\pm0.04$ \cite{Tsang01}. Sequential decays may modify the alpha values. However, results from current sequential decay calculations are model dependent \cite{Tsang06, Maria08}. Since the differences in $\alpha$ values are small, it will be difficult to determine the exact effects until more realistic sequential model calculations become available.
\begin{figure}[htbp]
\centering
\includegraphics[angle=270,scale=0.5]{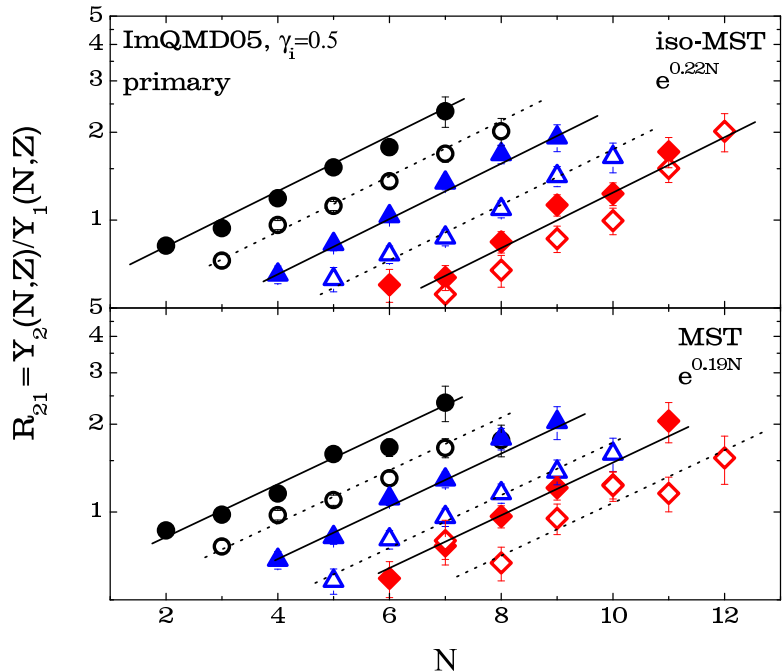}
\setlength{\abovecaptionskip}{50pt}
\caption{\label{ref-isoscaling-fig4} (Color online) $R_{21}$ values obtained from the ratios of the primary
isotopic distributions for $^{124}Sn+^{124}Sn$ collisions divided by those for $^{112}Sn+^{112}Sn$ collisions,
for $\gamma_i=0.5$ at b=2fm. The upper panel shows the results obtained with the iso-MST case for Z=3 to 8 and the lower panel shows the results obtained with the MST case for Z=3 to 8. The best fit isoscaling parameters are listed in the panels. }
\setlength{\belowcaptionskip}{0pt}
\end{figure}

Finally, we examine the effects of isospin cluster recognition on the system equilibrium measured by the production of light particles. To characterize the degree of equilibrium reached in the collision system, the observable of $vartl=\sigma_{trans}^2/\sigma_{long}^2$ is often used in experimental and theoretical studies \cite{Resid, Zhang06, Zhang07}. Here, the $\sigma_{tran}^2$ and $\sigma_{long}^2$ are the variance of transverse and longitude rapidity distribution of fragment yields. When the reaction system reaches equilibrium, the value of $\sigma_{trans}^2/\sigma_{long}^2$ is close to 1. In Fig.5, we plot the calculated results of longitudinal (left panel) and transverse (right panel) rapidity distribution for the Z-weighted yield (Z=1-6) for $^{124}Sn+^{124}Sn$ at b=2fm. The much wider longitudinal rapidity distributions suggest that the systems are far from equilibrium. More interestingly, the peaks in the longitudinal rapidity distribution (left pane) for Z=1-6 particles obtained in the MST case diminishes in the iso-MST case. It suggests that a higher equilibrium degree is predicted in the iso-MST case. With $\gamma_i=0.5$, the value of $vartl=\sigma_{trans}^2/\sigma_{long}^2$ obtained with the iso-MST case is 0.62, which is slightly larger than 0.58 for the MST case. Thus in addition to the isospin sensitive observables, the equilibrium or stopping power of the system also depends on the detailed description of cluster formation implemented in the transport models as well as on the mean field and the in-medium NN cross section.
\begin{figure}[htbp]
\centering
\includegraphics[angle=270,scale=0.5]{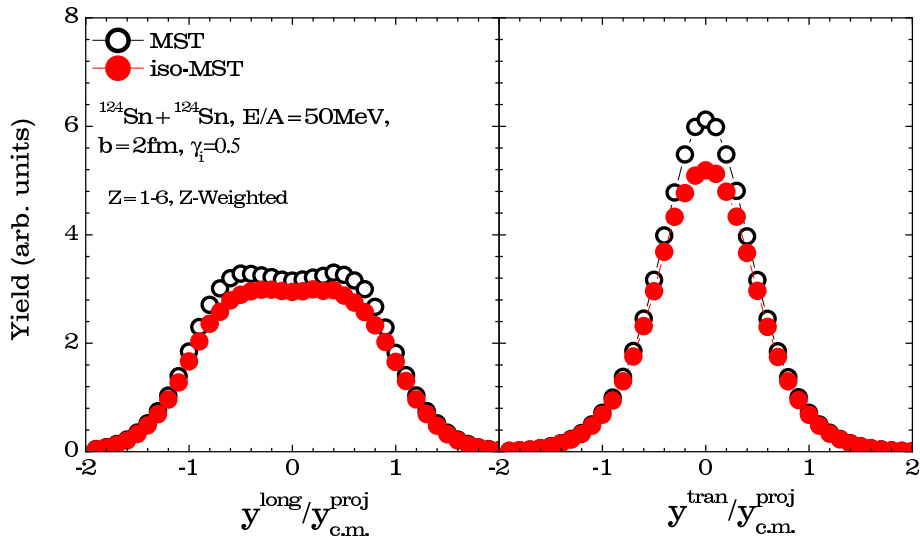}
\setlength{\abovecaptionskip}{50pt}
\caption{\label{ref-zweight-rapidity-fig5} (Color online) The calculated results of longitudinal (left panel)
and transverse (right panel) rapidity distribution for the Z-weighted yield (Z=1-6) for $^{124}Sn+^{124}Sn$ at
 b=2fm.}
\setlength{\belowcaptionskip}{0pt}
\end{figure}

In summary, we introduce a phenomenological isospin dependence in the description of cluster formation in transport models by adopting different $R_0$ values for pp, nn and np, $R_0^{pp}=3 fm$ and $R_0^{nn}=R_0^{np}=6 fm$. Our results using the isospin dependent minimum spanning tree method show suppression of Z=1 particles and enhancement of fragments, especially for heavier fragments with $Z\geq12$. Furthermore, we find enhanced production of neutron-rich isotopes at mid-rapidity. Consequently, isospin sensitive observables, such as the double ratios, $DR(t/^3He)$, and isoscaling parameter $\alpha$ increase to larger values. The widths of the longitudinal and transverse rapidity distributions of $Z=1-6$ particles also change. In all the observables that we examine, the effects introduced by the iso-MST algorithm are relatively small but in the direction of better agreement with data. However, we have not included sequential decays which may modify the magnitude of the effects. Nonetheless, the isospin dependence of the cluster recognition can be easily implemented and should be included in nuclear transport models.

\begin{acknowledgements}
This work has been supported by the Chinese National Science Foundation under Grants 11075215, 10979023, 10875031, 11005022,11005155, 10235030, and the national basic research program of China No. 2007CB209900. We wish to acknowledge the support of the National Science Foundation Grants No. PHY-0606007.
\end{acknowledgements}

\end{document}